\documentclass[runningheads]{llncs}
\usepackage[T1]{fontenc}
\usepackage[utf8]{inputenc}
\usepackage{graphicx}
\usepackage{fancyvrb}
\usepackage{xurl}
\usepackage{booktabs}
\usepackage{url}
\usepackage{amsmath,amsfonts}
\usepackage{algorithmic}
\usepackage{graphicx}
\usepackage{textcomp}
\usepackage{xspace}
\usepackage{subcaption}
\usepackage{tabularx}
\usepackage{amsmath}
\usepackage{balance}
\usepackage{framed}
\usepackage{chngpage}
\usepackage[many]{tcolorbox}
\usepackage{pgfplots}
\usepackage{siunitx}
\usepackage{balance}
\usepackage{enumitem}
\usepackage{multirow}

\usepackage{hyperref}

\usepackage{graphicx}
\usepackage{subcaption}

\newcommand{\claude}{\texttt{Claude.md}\xspace}
\newcommand{\am}{\texttt{agentic coding manifests}\xspace}
\newcommand{\Am}{\texttt{Agentic coding manifests}\xspace}
\newcommand{\AM}{\texttt{Agentic Coding Manifests}\xspace}
\newcommand{\ACMs}{\texttt{ACMs}\xspace}
\newcommand{\numfiles}{\texttt{253}\xspace}
\newcommand{\numrepos}{\texttt{242}\xspace}
\newcommand{\numcommits}{\texttt{1,249}\xspace}
\newcommand{\codex}{\texttt{AGENTS.md}\xspace}

\newcommand{\rqa}{$RQ_1$}

\newcommand{\rqc}{$RQ_2$}

\newcommand{\rqaa}{How Are Agent Manifest Files Organized?}

\newcommand{\rqcc}{What Instruction Are Included?}

\newcommand{\rqA}{\rqa: \rqaa}

\newcommand{\rqC}{\rqc: \rqcc}

\newcommand{\lblGenSysOverview}{\texttt{System Overview}\xspace}
\newcommand{\cntGenSysOverview}{48.2}

\newcommand{\lblClaudeAIIntegration}{\texttt{AI Integration}\xspace}
\newcommand{\cntClaudeAIIntegration}{15.4}

\newcommand{\cntDocRefs}{13.8}

\newcommand{\lblArchitecture}{\texttt{Architecture}\xspace}
\newcommand{\cntArchitecture}{64.8}

\newcommand{\lblImplDetails}{\texttt{Implementation Details}\xspace}
\newcommand{\cntImplDetails}{71.9}

\newcommand{\lblBuildRun}{\texttt{Build and Run}\xspace}
\newcommand{\cntBuildRun}{77.1}

\newcommand{\lblTest}{\texttt{Testing}\xspace}
\newcommand{\cntTest}{60.5}

\newcommand{\cntConfigEnv}{26.5}

\newcommand{\lblDeployOps}{\texttt{DevOps}\xspace}
\newcommand{\cntDeployOps}{9.5}

\newcommand{\lblMaintenance}{\texttt{Maintainance}\xspace}
\newcommand{\cntMaintenance}{19.8}

\newcommand{\lblProjMgmt}{\texttt{Project Management}\xspace}
\newcommand{\cntProjMgmt}{11.1}

\newcommand{\lblDevProcess}{\texttt{Development Process}\xspace}
\newcommand{\cntDevProcess}{37.2}

\newcommand{\lblPerformance}{\texttt{Performance}\xspace}
\newcommand{\cntPerformance}{12.7}

\newcommand{\lblSecurity}{\texttt{Security}\xspace}
\newcommand{\cntSecurity}{8.7}

\newcommand{\lblUIUX}{\texttt{UI/UX}\xspace}
\newcommand{\cntUIUX}{8.3}

\newcommand{\openAPRs}{54}
\newcommand{\mergedAPRs}{475}
\newcommand{\closedAPRs}{92}

\pgfmathsetmacro{\collectedAPRs}{int(\closedAPRs+\openAPRs+\mergedAPRs)}
\pgfmathsetmacro{\studiedAPRs}{int(\mergedAPRs+\closedAPRs)}

\pgfmathsetmacro{\calcMergeRateAPR}{(\mergedAPRs/\studiedAPRs)*100}

\pgfmathsetmacro{\calcRejectRateAPR}{1-\calcMergeRateAPR}

\newcommand{\directMergedAPRs}{208}
\pgfmathsetmacro{\calcDirectMergeRateAPR}{(\directMergedAPRs/\mergedAPRs)*100}
\pgfmathsetmacro{\calcInDirectMergeRateAPR}{(1-(\directMergedAPRs/\mergedAPRs))*100}

\newcommand{\mergedHPRs}{521}
\newcommand{\closedHPRs}{46}

\pgfmathsetmacro{\collectedHPRs}{int(\closedHPRs+\mergedHPRs)}
\pgfmathsetmacro{\studiedHPRs}{int(\mergedHPRs+\closedHPRs)}

\pgfmathsetmacro{\calcMergeRateHPR}{(\mergedHPRs/\studiedHPRs)*100}
\pgfmathsetmacro{\calcRejectRateHPR}{1-\calcMergeRateHPR}

\newcommand{\directMergedHPRs}{219}
\pgfmathsetmacro{\calcDirectMergeRateHPRs}{(\directMergedHPRs/\mergedHPRs)*100}

\definecolor{darkgreen}{rgb}{0, 0.5, 0} 
\definecolor{whitesmoke}{rgb}{0.99, 0.99, 0.99}

\def\Underline{\setbox0\hbox\bgroup\let\\\endUnderline}
\def\endUnderline{\vphantom{y}\egroup\smash{\underline{\box0}}\\}
\def\|{\verb|}

\newcommand{\ie}{\textit{i.e.,}\xspace}
\newcommand{\eg}{\textit{e.g.,}\xspace}

\newcommand{\etal}{\xspace\textit{et al.}\xspace}

\newcounter{findings_no}

\usepackage{listings}
\definecolor{backcolour}{rgb}{0.95,0.95,0.92}
\lstdefinelanguage{diff}{
  morecomment=**[f][\color{red}]{-},         
  morecomment=**[f][\color{darkgreen}]{+},       
  moredelim=**[is][\bfseries]{@@}{@@},
}
\definecolor{backcolour}{rgb}{0.95,0.95,0.92}
\lstdefinelanguage{commit}{ 
  breakindent = 0pt,
  numbers=none,
  backgroundcolor=\color{white},
  frame=single,
  xleftmargin=3.5em,
  numbersep=0em,
  xrightmargin=1.5em,
}

\usepackage[many]{tcolorbox}  
\tcbuselibrary{listings,breakable}
\definecolor{main}{HTML}{D0D3D4}    
\definecolor{sub}{HTML}{D0D3D4}     
\newtcolorbox{dbox}{
    left=2pt,right=2pt,top=2pt,bottom=2pt,
    enhanced, 
    boxrule = 0pt,
    enlarge top by=5pt,
    enlarge bottom by=3pt,
  }
\tcbset{
    sharp corners,
    before skip = 0.1cm,    
    after skip = 0.01cm      
}

\begin{document}
\title{On the Use of Agentic Coding Manifests: \\An Empirical Study of Claude Code}
\titlerunning{On the Use of Agentic Coding Manifests}
%
\author{
Worawalan Chatlatanagulchai\inst{1} \and
Kundjanasith Thonglek\inst{1} \and
Brittany Reid\inst{2} \and
Yutaro Kashiwa\inst{2} \and
Pattara Leelaprute\inst{1} \and
Arnon Rungsawang\inst{1} \and
Bundit Manaskasemsak\inst{1} \and
Hajimu Iida\inst{2}
}
%
\authorrunning{W. Chatlatanagulchai \etal}
%
\institute{Faculty of Engineering, Kasetsart University, Thailand
\and Nara Institute of Science and Technology (NAIST), Japan}
\maketitle              
\begin{abstract}
Agentic coding tools receive goals written in natural language as input, break them down into specific tasks, and write/execute the actual code with minimal human intervention. Key to this process are agent manifests, configuration files (such as \claude) that provide agents with essential project context, identity, and operational rules. However, the lack of comprehensive and accessible documentation for creating these manifests presents a significant challenge for developers. We analyzed \numfiles \claude files from \numrepos repositories to identify structural patterns and common content. Our findings show that manifests typically have shallow hierarchies with one main heading and several subsections, with content dominated by operational commands, technical implementation notes, and high-level architecture.
\end{abstract}
\section{Introduction}
\textit{``The hottest new programming language is English,''} stated OpenAI founding member Andrej Karpathy,\footnote{\url{https://x.com/karpathy/status/1617979122625712128}} capturing a fundamental shift in how software development is evolving. The rise of Large Language Models (LLMs) has enabled the deployment of Artificial Intelligence (AI) agents capable of facilitating or executing autonomous software engineering tasks through natural language interactions. This novel approach, termed \texttt{Agentic Coding}, interpret natural language goals, decompose them into subtasks, and autonomously plan and execute code with minimal human intervention. Unlike vibe coding, which focuses on describing desired feelings or essence, agentic coding provides concrete objectives and lets AI independently determine implementation through multi-step planning, tool usage, and self-correction.

Notable implementations of agentic coding tools include Claude Code, Cursor, Aider, GitHub Copilot, and Devin AI. Most of these tools rely on \Am to function effectively, which are specialized configuration files that define AI agent behavior within specific projects. Loaded at the start of each session, they equip AI agents with project-specific knowledge, behavioral guidelines, and operational rules that determine how well the agent can understand codebases, interpret developer intent, and execute tasks autonomously. However, despite \am being crucial for the performance, there is little research on how to design them effectively. The lack of comprehensive documentation means developers resort to trial-and-error approaches, resulting in suboptimal agent behavior and missed opportunities to fully leverage these tools' capabilities.

This study aims to reveal common structural patterns and their contents. Our empirical study on \numfiles \claude files from \numrepos repositories revealed (i) they follow a shallow hierarchical structure with a single main heading and moderate subsections; and (ii) the most common content patterns include instructions for \texttt{\lblBuildRun}, \texttt{\lblImplDetails}, and \texttt{\lblArchitecture}, highlighting the critical role of contextual information for AI-assisted software development and how action-oriented focus of \am.
\vspace{-2mm}\section{Agent Coding Manifests}\vspace{-2mm}
Agentic coding tools can be configured through specialized Markdown files that define how AI coding assistants should operate within specific projects. These configuration files, such as \claude for Claude Code or \codex for Codex, establish the AI agent's identity, capabilities, and operational workflows.\footnote{An example can be seen here: \url{https://github.com/fschutt/azul/blob/3fe83b9d4c8004ebe96ea0a77660c777fcd05bc8/CLAUDE.md}} 
We refer to these configuration files as \AM (\ACMs). By documenting project-specific context and conventions, \ACMs eliminate the need for repetitive explanations and reduces misunderstandings between developers and AI assistants. When stored in version control alongside the codebase, these files ensure that AI agents maintain a consistent understanding of each project's unique requirements and characteristics throughout the development lifecycle.

There are official documents for each agentic coding tool that introduce setup for \ACMs to pull context into prompts automatically. However, it is not comprehensively written.  For example, the official documents of Claude Code indicate only that \ACMs can be used to share instructions for the project, such as project architecture, coding standards, and common workflows. 

This fragmentation and lack of explicit, standardized guidance can delay developers' ability to effectively define, orchestrate, and leverage the Claude agent's behavior, therefore creating a substantial barrier to exploiting the full potential of agentic coding and potentially leading to inconsistencies in agent performance and a steeper learning curve for integration. Consequently, this gap in practical guidance for the creation of \ACMs served as a primary motivation for our current research. We aim to address this challenge by systematically investigating existing \claude files within open-source repositories. Our study is specifically designed to infer common structural patterns, typical instructions, and general practices in how developers configure these crucial \am and maintain them.


\section{Data Collection}

This study used a systematic data collection methodology using the GitHub API\footnote{\url{https://docs.github.com/en/rest?apiVersion=2022-11-28}} to identify and analyze ``\claude'' files in open-source software repositories. 
The data collection process began by searching for repositories containing files named \claude (insensitive to case) using the GitHub API. The search focused on files created between February 24, 2025, when Claude MCP\footnote{https://docs.anthropic.com/en/release-notes/claude-code} was released, and June 16, 2025. This initial search identified 838 \claude files distributed in 806 different repositories. 

To exclude projects that had only recently adopted Claude Code, we applied a filtering requiring repositories to have at least 20 commits after introducing their \claude file. 
This threshold corresponds to an average of five commits per month during the four months between the release of Claude Code and the start of data collection. After applying this filter, our dataset comprised  \numfiles \claude files from \numrepos repositories. We cloned these repositories and retrieved \numcommits commits in total.





\section{Results}
\subsection*{\rqA}

\textbf{Motivation.}
Prior work on software documentation has noted that developer docs often use hierarchical section structures~\cite{7000568}. For example, Treude\etal observed that technical documentation ``usually follows a hierarchical structure with sections and subsections'' \cite{7000568}. Empirical studies of project documentation (\eg \texttt{README} and \texttt{CONTRIBUTING} files) also find that early versions tend to be very minimal and focused on basic usage or contribution procedures~\cite{gaughan2025introductionreadmecontributingfiles}. However,  \am files such as \texttt{CLAUDE.md} represent a novel documentation artifact specifically designed for AI-assisted coding. To the best of our knowledge, no prior research has systematically analyzed their structural characteristics (\eg how many and what levels of instructions they contain). We first investigate common organizational patterns that developers use when structuring instructions for AI coding agents.

\textbf{Approach.}
We extracted each \claude file from the cloned repositories and measured the number of markdown headers that multiply nest. Specifically, we identified each header level, ranging from H1 (\ie \texttt{\#}) to H6 (\ie \texttt{\#\#\#\#\#\#}). A header's section includes all subsequent lines until the next header of the same level is found. We counted every non-empty line within these sections. Lines within code blocks (\ie demarcated by \texttt{\`{}\`{}\`{}}) were not included in our count. 

\begin{figure}[t]
    \centering
    \includegraphics[width=\textwidth]{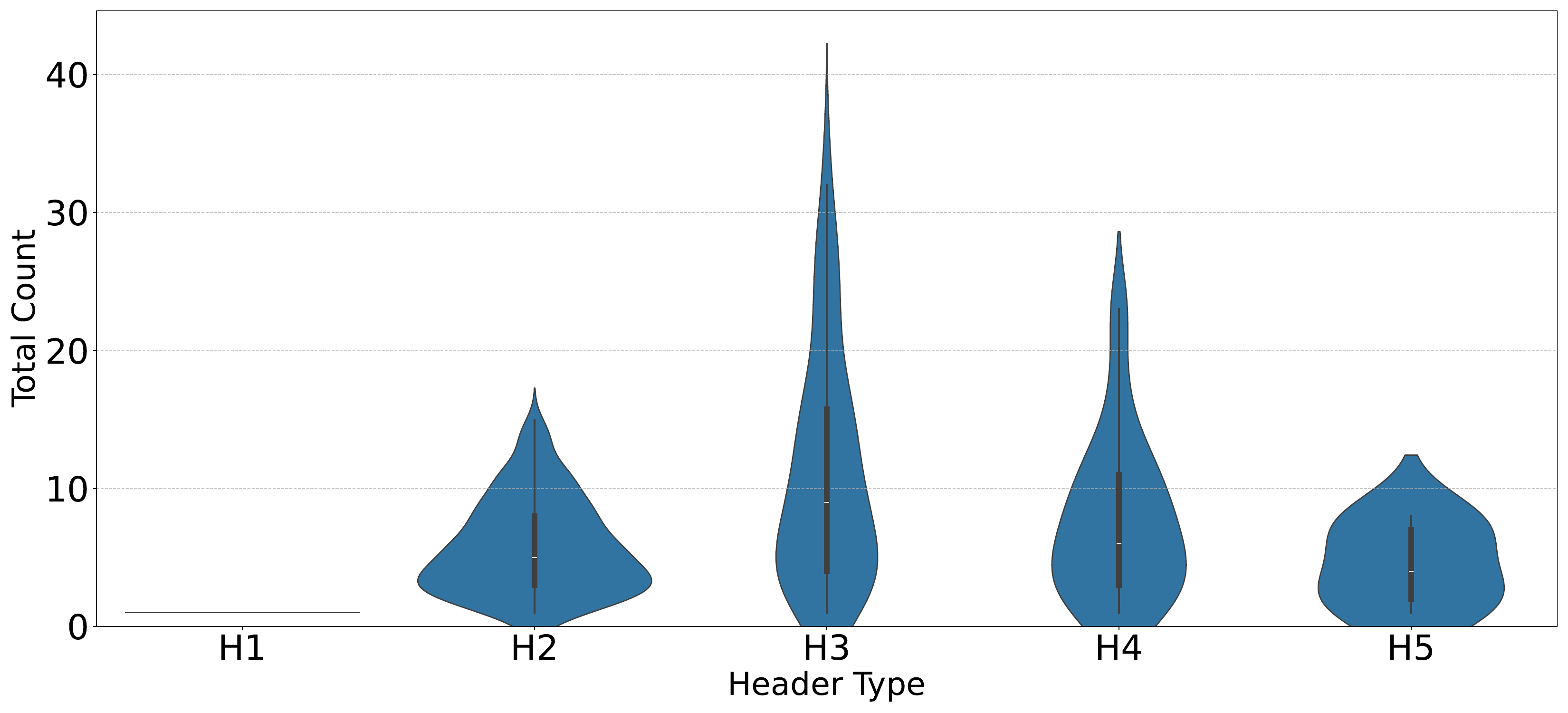}
    \caption{Distribution of header counts across \claude files (outliers removed)}
    \label{fig:total_headers_violin_plot}
\end{figure}
\textbf{Results.}
Figure~\ref{fig:total_headers_violin_plot} depicts the distribution of header levels in \claude files. 
The distributions indicate that most documents begin with a single primary heading (H1), with a median of 1.0. This typically branches into a moderate number of subsections (H2), showing a median of 5.0, followed by more granular points (H3), which have a median of 9.0.

Header usage declines sharply as depth increases. Deeper levels such as H4 and H5 occur infrequently, with H4 appearing in only 37 documents and H5 in just 5 among the \numfiles\ files analyzed. The median counts per document are 6 for H4 and 4 for H5, which indicates their limited use. We observed H6 only once in the entire corpus; given this extremely low frequency, we exclude it from further analysis. Overall, deeply nested structures are rare, indicating a preference for straightforward organization. 

This organizational pattern in \am is further confirmed by the statistics for nested headers, which show a similar distribution to the overall heading counts. For example, the median for H2 headings directly nested under an H1 heading (Median = 5.0) is nearly identical to the overall statistics for all H2 headings (Median = 5.0). This consistency suggests a predictable structural approach across the dataset, where the main topics are broken down in a similar fashion. We observed distinct purposes for different heading levels. H1 headings typically encapsulate the main content topic of the \am. H2 headings commonly describe broader aspects such as coding style, project structure, command-line instructions, or overall testing strategies. As the hierarchy deepens, H3 headings become more granular, detailing specific methods of testing or how to apply those methods. Further details on the content will be addressed in the next research question.

\smallskip
\begin{tcolorbox}
\textbf{Answer to RQ1:}
Agent manifest files are typically organized using a shallow hierarchical structure, with most documents starting from a single top-level heading and branching into a moderate number of H2 and H3 subsections.
\end{tcolorbox}



\subsection*{\rqC}

\ \ \ \ \textbf{Motivation.}
Prior work shows that clear, structured instructions, such as stepwise task descriptions or templated formats, significantly improve LLM-generated outputs \cite{zamfirescu2023prompting}, with performance fundamentally shaped by the contextual information provided \cite{min2022rethinking}. Despite this, there is little empirical research on \am intended to configure and guide AI agents. RQ2 addresses this by identifying prevalent instruction patterns in these manifests, revealing how developers structure context to align AI behavior in practice.

\smallskip
\textbf{Approach.}
We adopted a two-stage manual content classification approach comprising a label creation phase followed by a label assignment phase. This separation was necessary due to the extensive structure and diversity of instructional content in \claude, which made simultaneous label generation and assignment impractical.

In the first phase, we focused on constructing a robust and comprehensive label set. We began by extracting all the H1 and H2 titles from the \claude files. Subsequently, we prompted three popular large language models (LLMs), Claude, Gemini, and ChatGPT, to generate candidate labels. One of the authors then selected the most appropriate label from these suggestions or created a new label when none were suitable. The use of LLMs was motivated by findings from prior research~\cite{DBLP:conf/msr/AhmedDTP25}, which demonstrated that recent LLMs perform comparably to human annotators in manual labeling tasks while significantly reducing effort.
To ensure label quality, two authors independently reviewed the initial label set. This process yielded 80 distinct labels. In the final step, three inspectors collaboratively refined the label set by merging semantically similar entries, resulting in a consolidated set of 20 core labels.

In the second phase, two inspectors assigned the labels generated in the first phase to each \claude file, allowing multiple labels per file. Initially, both inspectors independently labeled the content of each file. This process resulted in 1,228 total label assignments across the \numfiles files, with 113 instances of disagreement. To resolve these conflicts, a third inspector joined the discussion and collaborated with the initial two to reach a consensus on the final labels. During this reconciliation process, a new and more descriptive label (\ie \lblBuildRun) was introduced. This refinement finalized our classification scheme with 15 distinct labels. All three inspectors involved in the labeling process have programming experience ranging from 4 to 17 years.

\begin{table}[t]
\scriptsize
    \centering
    \caption{Categories, descriptions, and their prevalence.}
    \begin{tabular}{clp{7cm}r}
        \toprule
        Category & Label & Description & \% \\
        \midrule
        General & \lblGenSysOverview & Provides a general overview or describes the key features of the system. & \cntGenSysOverview \\\addlinespace[1mm] 
        & \lblClaudeAIIntegration & Contains instructions or notes specifically for integrating with or interacting with the agentic coding tools. & ~\cntClaudeAIIntegration \\\addlinespace[1mm] 
        & \texttt{Doc.\&Refs} & Lists supplementary documents, links, or references for additional context. & \cntDocRefs \\\midrule
        Impl. & \lblArchitecture & Describes the high-level structure, design principles, or key components of the system's architecture. & \cntArchitecture \\\addlinespace[1mm] 
        & \texttt{Impl. Details} & Provides specific details for implementing code or system components, including coding style guidelines. & \cntImplDetails \\\midrule
        Build & \lblBuildRun & Outlines the process for compiling source code and running the application, often including key commands. & \cntBuildRun \\\addlinespace[1mm] 
        & \lblTest & Details the procedures and commands for executing automated tests. & \cntTest \\\addlinespace[1mm] 
        & \texttt{Conf.\&Env.} & Instructions for configuring the system and setting up the development or production environment. & \cntConfigEnv \\\addlinespace[1mm] 
        & \lblDeployOps & Covers procedures for software deployment, release, and operations, such as CI/CD pipelines. & \cntDeployOps \\\midrule
        Management         & \lblDevProcess & Defines the development workflow, including guidelines for version control systems like Git. & \cntDevProcess \\\addlinespace[1mm] 
& \lblProjMgmt & Information related to the planning, organization, and management of the project. & \cntProjMgmt 
\\\midrule
        Quality & \lblMaintenance & Guidelines for system maintenance, including strategies for improving readability, detecting and resolving bugs. & \cntMaintenance \\\addlinespace[1mm] 
        & \lblPerformance & Focuses on system performance, quality assurance, and potential optimizations. & \cntPerformance \\\addlinespace[1mm] 
        & \lblSecurity & Addresses security considerations, vulnerabilities, or best practices for the system. & \cntSecurity \\\addlinespace[1mm] 
        & \lblUIUX & Contains guidelines or details concerning the user interface (UI) and user experience (UX). & \cntUIUX \\
        \bottomrule
    \end{tabular}
    \label{tab:category_summary}
\end{table}

\smallskip
\textbf{Results.}
Table~\ref{tab:category_summary} presents the distribution of documentation categories, where percentages indicate the proportion of \claude files containing instructions for each category.
The most prevalent was \lblBuildRun (\cntBuildRun\%), containing command-line instructions, scripts, and procedures for compiling and running code. This was followed by \lblImplDetails (\cntImplDetails\%) with development guidance (\eg code style) and \lblArchitecture (\cntArchitecture\%) describing high-level system design.

While the most prevalent instructions (\lblBuildRun, \lblImplDetails, \lblArchitecture, \lblTest) address functional aspects, meta-level or non-functional categories like Performance (\cntPerformance\%), Security (\cntSecurity\%), and UI/UX (\cntUIUX\%) appear far less frequently. This pattern suggests that manifests are primarily optimized to help agents execute and maintain code efficiently rather than address broader quality attributes or user-facing aspects.

Beyond functional factors, we observed notable instances where developers provide contextual information. For example, half of the \claude files contain system overview explanations (\ie \lblGenSysOverview). Additionally, \cntClaudeAIIntegration\% of manifests (\ie \lblClaudeAIIntegration label) explicitly define the agent's role and describe its responsibilities within the project (\eg reviewers). This indicates that manifests serve not only as technical guides but also as means of establishing an AI agent's understanding, responsibilities, and collaborative alignment.

\smallskip
\begin{tcolorbox}
\textbf{Answer to RQ2:}
The most common content categories in \am are \lblBuildRun, followed by \lblImplDetails and \lblArchitecture   descriptions. These patterns reflect the action-oriented focus and specificity of the files.
\end{tcolorbox}



    
\section{Future Direction}\vspace{-2mm}
Our future work will pursue the following two directions as well as involving more different agentic coding tools, such as Codex and Copilot.

\textbf{Maintenance:}
The long-term efficacy and relevance of \am may be inherently tied to their maintenance. Previous research on documentation maintenance, such as Gaughan\etal\cite{gaughan2025introductionreadmecontributingfiles} observed ``burst-then-taper'' effect in documentation changes. Future work could examine the evolution and decay of \claude files, revealing update frequencies, change-prone sections, and correlations with project development cycles or agent performance. 

\textbf{Impact:}
A critical direction for future research is to empirically assess the direct impact of \claude files on the performance of the Claude agent and the productivity of developers utilizing it. This could involve controlled experiments where agents are tasked with identical coding challenges, but with varying qualities or completeness of \am. Metrics such as task completion time, code quality (\eg bug count, adherence to style guides), number of iterations, and developer satisfaction could be measured. Additionally, qualitative studies, such as developer interviews or surveys, could explore how developers perceive the utility and influence of well-crafted manifests on their workflow, debugging efforts, and overall experience with agentic coding. 

\vspace{-3mm}\section{Related Work}\vspace{-2mm}
\ \ \ \ \ \textbf{Instructions for AI:}
A growing body of work investigates how developers formulate instructions for AI tools (prompt engineering) and categorizes AI agent tasks. Schulhoff\etal\cite{schulhoff2025promptreportsystematicsurvey} investigates a wide array of prompt techniques and best practices. Kumar\etal\cite{kumar2025sharptoolsdeveloperswield} examine recurring instruction patterns and found that roughly 50\% of developer prompts to an AI agent requested code changes, while others sought code explanations, test execution, or reviews. These studies examine typical instruction patterns that directly order AIs, while our study focuses on the contexts written in documents behind the instructions.

\textbf{Contexts for AI:}
Many studies~\cite{brezillon_context_ai_II,calegario2023exploringintersectiongenerativeai} consistently highlight the critical role of contextual information such as codebase~\cite{Athale_2025}, documentation~\cite{wang2025understandingcontextutilizationcode}, and dependencies~\cite{hai2025impactscontextsrepositorylevelcode} for effective AI software development assistance.  Akhoroz\etal\cite{akhoroz2025conversationalaicodingassistant} observed that programmers often note ``\textit{inaccuracies [and] lack of contextual awareness}'' in AI outputs. Tufano\etal~\cite{tufano2024autodevautomatedaidrivendevelopment} points out that existing assistants like Copilot ``\textit{exhibit limited functionalities and lack contextual awareness}''. 
Our work directly addresses this need, as \am are designed to encapsulate this crucial context.

\textbf{Documents for Human:}
Previous studies provide a strong foundation for understanding the structure and evolution of documentation. Gaughan \etal\cite{gaughan2025introductionreadmecontributingfiles} found that initial README files are typically concise and function-focused, often expanding over time. Similarly, Prana\etal \cite{prana2019categorizing} observed that over 90\% of GitHub READMEs they studied mentioned basic information like project name, description, and usage instructions. Our work extends this by quantifying the structure of the documents for AIs, like \claude.

\newpage
\section{Conclusion}\vspace{-2mm}
This study analyzed \numfiles \claude files and found that developers prefer a shallow hierarchical structure, typically using a single primary heading with a moderate number of subsections. Additionally, manual analysis revealed that manifests prioritize operational commands, followed by technical implementation notes and architecture descriptions. Many manifests also include AI role definitions to guide agent behavior. These patterns demonstrate how manifests serve dual purposes in agentic coding workflows: as execution guides for technical tasks and as frameworks for human-AI collaboration.

\vspace{-4mm}
\subsection*{\ackname}\vspace{-2mm}
We gratefully acknowledge the financial support of JSPS KAKENHI grants (JP24K02921, JP25K21359), as well as JST PRESTO grant (JPMJPR22P3), ASPIRE grant (JPMJAP2415), and AIP Accelerated Program (JPMJCR25U7).

\vspace{-3mm}
\bibliographystyle{splncs04}
\bibliography{references.bib}

\end{document}